# Atmosphere-Ionosphere Response to the M9 Tohoku Earthquake Revealed by Joined Satellite and Ground Observations. Preliminary results.

Dimitar Ouzounov[1,2], Sergey Pulinets[3,5], Alexey Romanov[4], Alexander Romanov[4], Konstantin Tsybulya[3], Dimitri Davidenko[3], Menas Kafatos[1] and Patrick Taylor [2]

[1] Chapman University, One University Drive, Orange, CA 92866, USA

[2] NASA Goddard Space Flight Center, Greenbelt, MD 20771, USA

[3] Institute of Applied Geophysics, Rostokinskaya str., 9, Moscow, 129128, Russia

[4] Russian Space Systems, 53 Aviamotornya Str, Moscow, 111250, Russia

[5] Space Research Institute RAS, Profsoyuznaya str. 84/32, Moscow 117997, Russia

Correspondence to: D.Ouzounov(Ouzounov@chapman.edu; Dimitar.P.Ouzounov@nasa.gov)

**Abstrac**t

The recent M9 Tohoku Japan earthquake of March 11, 2011 was the largest recorded earthquake ever to hit this nation. We retrospectively analyzed the temporal and spatial variations of four different physical parameters - outgoing long wave radiation (OLR), GPS/TEC, Low-Earth orbit tomography and critical frequency foF2. These changes characterize the state of the atmosphere and ionosphere several days before the onset of this earthquake. Our first results show that on March 8[th] a rapid increase of emitted infrared radiation was observed from the satellite data and an anomaly developed near the epicenter. The GPS/TEC data indicate an increase and variation in electron density reaching a maximum value on March 8. Starting on this day in the lower ionospheric there was also confirmed an abnormal TEC variation over the epicenter. From March 3-11 a large increase in electron concentration was recorded at all four Japanese ground based ionosondes, which return to normal after the main earthquake. We found a positive correlation between the atmospheric and ionospheric anomalies and the Tohoku earthquake. This study may lead to a better understanding of the response of the atmosphere /ionosphere to the Great Tohoku earthquake.

## 1. Introduction

The 11 of March earthquake triggered was followed by a large number of powerful aftershocks. The possibility of a mega-earthquake in Miyagi prefecture was initially discussed by Kanamori et al. (2006). Strong earthquakes in this region were recorded since 1793 with average period of 37 ± 7 years. The latest great Tohoku earthquake matched this reoccurrence period since the last one occurred in 1978.

 The observational evidence, from the last twenty years, provides a significant pattern of transient anomalies preceding earthquakes (Tronin et al., 2002; Liu et al., 2004; Pulinets and Boyarchuk, 2004; Tramutoli et al., 2004, Parrot 2009, Oyama 2011). Several indicate that atmospheric variability was also detected prior to an earthquake. Despite these pre-earthquake atmospheric transient phenomenon (Ouzounov et al., 2007; Inan et al., 2008; Němec et al., 2009; Kon et al., 2011), there is still lack of consistent data necessary to understanding the connection between atmospheric and ionospheric associated with major earthquakes. In this present report we analyzed ground and satellite data to study the relationship between the atmospheric and ionospheric and the March 11 Tohoku earthquake.

We examined four different physical parameters characterizing the state of the atmosphere/ionosphere during the periods before and after the event: 1. Outgoing Longwave Radiation, OLR (infra-red 10-13 µm) measured at the top of the atmosphere; 2. GPS/TEC (Total Electron Content) ionospheric variability; 3. Low Earth Orbiting (LEO) satellite ionospheric tomography; and 4. Variations in ionosphere F2 layer at the critical foF2 frequency (the highest frequency at which the ionospheric is transparent) from four Japanese ionosonde stations. These multidisciplinary data provide a synopsis of the atmospheric/ionospheric variations related to tectonic activity.

## 2. Data Observation and Analysis
### 2.1 Earth radiation observation

One of the main parameters we used to characterize the earth's radiation environment is the outgoing long-wave-earth radiation (OLR). OLR has been associated with the top of the atmosphere integrating the emissions from the ground, lower atmosphere and clouds

(Ohring, G. and Gruber, 1982) and primary been used to study Earth radiative budget and climate (Gruber, A. and Krueger, 1984; Mehta, A., and J. Susskind, 1999)

The National Oceanic and Atmospheric Administration (NOAA) Climate Prediction Center (http://www.cdc.noaa.gov/) provides daily and monthly OLR data. The OLR algorithm for analyzing the Advanced Very High Resolution Radiometer (AVHRR) data that integrates the IR measurements between 10 and 13 μm. OLR is not directly measured, but is calculated from the raw data using a separate algorithm (Gruber and Krueger, 1984). These data are mainly sensitive to near surface and cloud temperatures. A daily mean covering a significant area of the Earth (90° N- 90° S, 0° E to 357.5° E) and with a spatial resolution of 2.5° x2.5° was used to study the OLR variability in the zone of earthquake activity (Liu, 2000; Ouzounov et all, 2007, Xiong at al, 2010). An increase in radiation and a transient change in OLR were proposed to be related to thermodynamic processes in the atmosphere over seismically active regions. An anomalous eddy of this was defined by us (Ouzounov et al, 2007) as an E_index. This index was constructed similarly to the definition of anomalous thermal field proposed by (Tramutoli et al., 1999). The E_index represents the statically defined maximum change in the rate of OLR for a specific spatial locations and predefined times:

$$\Delta E\_Index(t) = (S^*(x_{i,j}, y_{i,j}, t) - \overline{S}^*(x_{i,j}, y_{i,j}, t))/\tau_{i,j} \qquad (1)$$

Where: t=1, K – time in days, $S^*(x_{i,j}, y_{i,j}, t)$ the current OLR value and $\overline{S}^*(x_{i,j}, y_{i,j}, t)$ the computed mean of the field, defined by multiple years of observations over the same location, local time and normalized by the standard deviation $\tau_{i,j}$.

In this study we analyzed NOAA/AVHRR OLR data between 2004 and 2011. The OLR reference field was computed for March 1 to 31 using all available data (2004-2011) and using a ±2 sigma confidence level (Fig.2). During 8-11 March, a strong transient OLR anomalous field was observed near the epicentral area and over the major faults, with a confident level greater than +2 sigma (Fig. 3). The first indication of the formation of a transient atmospheric anomaly was detected on March 8th three days before the Tohoku earthquake with a confidence level of 2 sigma above the historical mean value. The location of the OLR maximum value on March 11, recorded at 06.30 LT was collocated exactly with the epicenter. This rapid enhancement of radiation could be explained by an anomalous flux of the latent heat over the area of increased tectonic activity. Similar observations were observed within a few days prior to the most recent major earthquakes China (M7.9, 2008), Italy (M6.3, 2009), Samoa (M7, 2009), Haiti (M7.0, 2010) and Chile (M8.8, 2010) (Pulinets and Ouzounov, 2011, Ouzounov et al, 2011a,b).

**2.2 Ionospheric observation**

The ionospheric variability around the time of the March 11 earthquake were recorded by three independent techniques: the GPS TEC in the form of Global Ionosphere Maps (GIM) maps, ionospheric tomography, using the signal from low-Earth orbiting satellites (COSMOS), and data from the ground based vertical sounding network in Japan. The period of this earthquake was very environmentally noisy for our analysis since two (small and moderate) geomagnetic storms took place on the first and eleventh of March respectively (Fig 4B). There was a short period of quiet geomagnetic activity between March fifth and tenth but it was during a period of increasing solar activity. During period from 26 February through 8 March the solar F10.7 radio flux increased almost two-fold (from 88 to 155). So the identification of the ionospheric precursor was the search a signal in this noise.

To reduce this noise we used the following criteria:
1. If an anomaly is connected with the earthquake, it should be local (connected with the future epicenter position) contrary to the magnetic storms and solar activity that affected the ionosphere, which are global events.
2. All anomalous variation (possible ionospheric precursor) should be present in the records of all three ionosphere-monitoring techniques used in our analysis
3. The independent techniques concerning geomagnetic activity that were previously developed (Pulinets et al. 2004) were used.

The only source where we were able to get three spatially coincident anomalies was GPS GIM. We made four types of analysis: a.) Differential maps; b.) Global Electron Content (GEC) calculations (Afraimovich et al., 2008); c.) Determination of ionospheric anomaly local character; and d.) Variation of GPS TEC in the IONEX grid point (Pulinets et al. 2004) closest to the Tohoku earthquake epicenter.

To estimate variability of the GIM a map using the average of the previous 15 days, before March 11, was calculated and the difference DTEC between the two TEC maps was obtained by subtracting the current GIM from the 15-day average map. This value was selected at 0600 UT corresponding to 15.5 LT, when the equatorial anomaly is close to a maximum (one might expect the strongest variations at this local time). The most remarkable property of the differential maps was the sharp TEC increase during the recovery phase during March 5 through 8 where the strongest deviation from the average was recorded on 8 of March. This distribution is shown in Fig. 4 A. To understand if this increase was a result of the abrupt increase in solar activity and has a either local or global character we calculated the Global Electron Content (GEC) according to Afraimovich et al. (2008). In Fig. 5 one can see the solar F10.7 index variation (green) in comparison with GEC (blue). Both parameters were normalized to see their similarity. It is interesting to note that on the increasing phase both parameters are very close, the recovery phase shows the difference (2-3 days of ionosphere reaction delay in comparison with F10.7 (what corresponds to conclusions of Afraimovich et al (2008)). Two small peaks on the ionospheric (blue) curve on 1 and 13 March correspond to two small geomagnetic storms (see, Dst index in Fig. 4 B). To determine if there is any local anomaly in the region near the epicenter we integrated the GEC, but in a circular area with a radius of 30° around the epicenter. The normalized curve (with the same scale as the first two is given in red. And immediately one can observe the remarkable peak on 8 March. This date, March 8, corresponds to the day of the differential GIM shown in Fig 4 A. The local character of the ionospheric anomaly on has been demonstrated by this test.

This last check was made by studying the TEC variation at the grid point closest to the epicenter as shown in Fig 4 B (upper panel). One should keep in mind that only data for 0600 UT were taken, so we have only one point for this day. Again a strong and very unusual increase of TEC was registered on March 8 marked by red in figure 4. The effect of magnetic storms is marked in blue in this figure. Note the gradual trend of background TEC values, which is probably, connected with the general electron density increase at the equinox transition period (passing from winter to summer electron concentration distribution). From point measurements we observe that the most anomalous day is March 8.

The data used to derive an image of the base of the ionosphere tomography (Fig. 1, Fig.6) was obtained from the coherent receivers chain on the Sakhalin island (Russia). Computing the base of ionosphere tomography utilizes the phase-difference method, (Kunitsyn and Tereshchenko, 2001) which is contained in the applied tomography software (Romanov et al, 2009). A coherent phase difference of 150 and 400 MHz was used to measure the relative ionosphere TEC values. The source signals are from COSMOS - 2414 series, OSCAR-31 series and RADCAL, low-Earth orbiting satellites with near-polar orbital inclinations. The ionosphere irregularity was observed from the relatively slanting TEC variations (increasing to 1.5 TECU above background) and in the ionosphere electron concentration tomography reconstruction. These data from the Tuzhno-Sakhalinsk and Poronajsk receivers and DMSP F15 satellite signals (maximum elevation angle was 70º) were used for calculating ionospheric tomography. A tomography image anomaly was located at 45-46-north latitude deg. It extends some 100-150 km along latitude 45N and has a density that is 50% higher than background. The structure of the March 8 2011, 19:29 UTC ionospheric F2 layer was located by the significant anomalous electron concentration anomaly recorded from a series of reconstructions of the ionospheric tomography (Fig. 6A). The strength and position of the detected anomaly can be estimated from Fig. 6 B. It should be noted that as in the case of GIM maps analysis of the most anomalous ionospheric tomography was recorded on March 8[th]. The results of ionospheric tomography confirm the conclusion of our previous analysis concerning March 8 as an anomalous day.

Data from the four Japanese ground based ionosondes (location shown in the Fig. 1) were analyzed. All stations indicated a sharp increase in the concentration of electrons at the beginning of March, but as it was demonstrated by GIM analysis that this increase is most probably due to the increase in solar activity. It was shown by Pulinets et al. (2004) by cross-correlation analysis of daily variation with the critical frequency (or vertical TEC) could reveal ionospheric precursors even in presence of a geomagnetic disturbances. It explained the fact that ionospheric variations connected with the solar and geomagnetic disturbances (in case when the stations are in similar geophysical conditions and not too far one from another) are very similar with a cross correlation coefficient greater than 0.9. At the same time (taking into account the physical mechanism of seismo-ionospheric disturbances (Pulinets and Boyarchuk, 2004) ionospheric variations registered by station closest to an epicenter would be different from ones recorded by more distant receivers. The pair of stations Kokubunji-Yamagawa is most appropriate for such an analysis. Kokubunji is the closest station to the earthquake epicenter, and the latitudinal difference between Kokubunji and Yamogawa is not so significant that we can neglect the latitudinal gradient (Fig. 1). Pulinets et al. (2004) demonstrated that the cross-correlation coefficient for a pair of stations with differing distances to an earthquake epicenter drops a few days before the earthquake. In Fig. 7 the cross-correlation coefficient shows the maximum drop on March 8. From ground based ionospheric sounding data we received confirmation that March 8 was an anomalous day and the ionospheric variations probably connected with the earthquake process. Our results show that on March 8 three independent methods of the ionosphere monitoring were anomalous and ionospheric variations registered on this day were related to the Tohoku earthquake.

## 3. Summary and Conclusions

The joint analysis of atmospheric and ionospheric parameters during the M9 Tohoku earthquake has demonstrated the presence of correlated variations of ionospheric anomalies implying their connection with before the earthquake. One of the possible explanations for this relationship is the Lithosphere- Atmosphere- Ionosphere Coupling mechanism (Pulinets and Boyarchuk, 2004; Pulinets and Ouzounov, 2011), which provides the physical links between the different geochemical, atmospheric and ionospheric variations and tectonic activity. Briefly, the primary process is the ionization of the air produced by an increased emanation of radon (and other gases) from the Earth's crust in the vicinity of active fault (Toutain and Baubron, 1998; Omori et al., 2007; Ondoh, 2009). The increased radon emanation launches the chain of physical processes, which leads to changes in the conductivity of the air and a latent heat release (increasing air temperature) due to water molecules attachment (condensation) to ions (Pulinets et al., 2007; Cervone et al., 2006; Prasad et al., 2005). Our results show evidence that process is related to the Tohoku earthquakes March 8 through 11 with a thermal build up near the epicentral area (Fig 2 and Fig.3). The ionosphere immediately reacts to these changes in the electric properties of the ground layer measured by GPS/TEC over the epicenter area, which have been confirmed as spatially localized increase in the DTEC on March 8 (Fig.4A). The TEC anomalous signals were registered between two minor and moderate geomagnetic storms but the major increase of DTEC, on 8 March, was registered during a geomagnetically quiet period (Fig.4A, B). A sharp growth in the electron concentration for Japanese ionospheric stations (Fig. 7) were observed with maximum on March 8 and then returned to normal a few days after the main earthquake of March 11.

Our preliminary results from recording atmospheric and ionospheric conditions during the M9 Tohoku Earthquake using four independent techniques: (i) OLR monitoring on the top of the atmosphere; (ii) GIM- GPS/TEC maps; (iii) Low-Earth orbit satellite ionospheric tomography; and (iv) Ground based vertical ionospheric sounding shows the presence of anomalies in the atmosphere, and ionosphere occurring consistently over region of maximum stress near the Tohoku earthquake epicenter. These results do not appear to be of meteorological or related to magnetic activity, due to their long duration over the Sendai region. Our initial results suggest the existence of an atmosphere/ionosphere response triggered by the coupling processes between lithosphere, atmosphere and ionosphere preceding the M9 Tohoku earthquake of March 11, 2011

**Acknowledgments**

We wish to thank to NASA Godard Space Flight Center, Chapman University and European Framework program #7 project PRE-EARTHQUAKE for their kind support. We also thank NOAA/ National Weather Service National Centers for Environmental Prediction Climate Prediction Center for providing OLR data. The IONEX data in this study were acquired as part of NASA's Earth Science Data Systems and archived and distributed by the Crustal Dynamics Data Information System (CDDIS). The F10.7 data were acquired from NOAA Space Weather Prediction center. World Data Center (WDC), Geomagnetism, in Kyoto, Japan, provided the Dst index and the Kp indices.

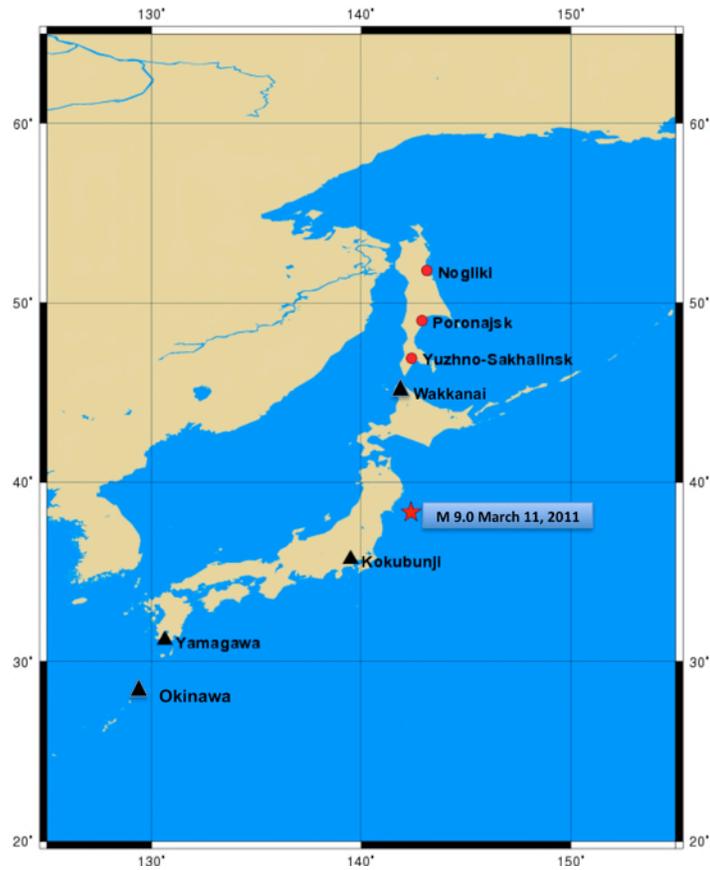

Figure 1. Reference map of Japan with the location of the M9.0 Tohoku Earthquake, March 11, 2011 (with red star). With red circles showing the location of the tomographic data receivers and with black triangles the location of vertical ionosonde stations in Japan.

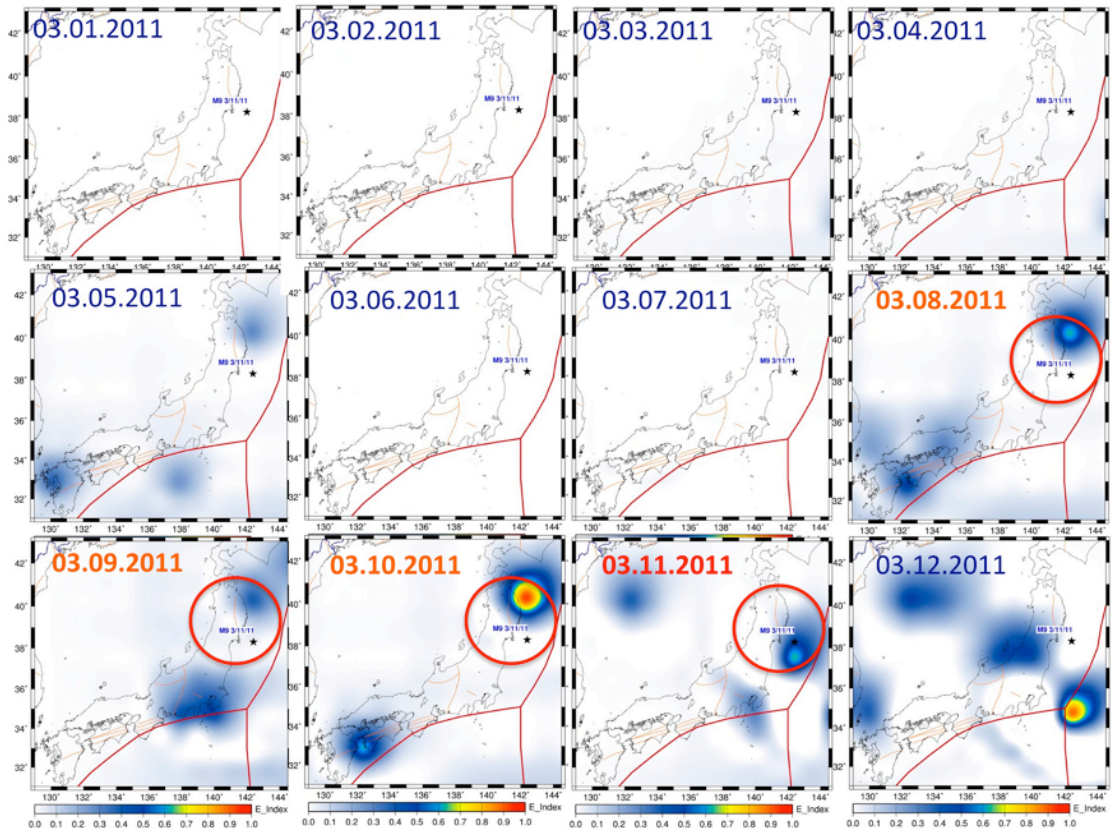

Figure 2. Time series of daytime anomalous OLR observed from NOAA/AVHRR (06.30LT equatorial crossing time) March 1-March12, 2011. Tectonic plate boundaries are indicated with red lines and major faults by brown ones and earthquake location by black stars. Red circle show the spatial location of abnormal OLR anomalies within vicinity of M9.0 Tohoku earthquake.

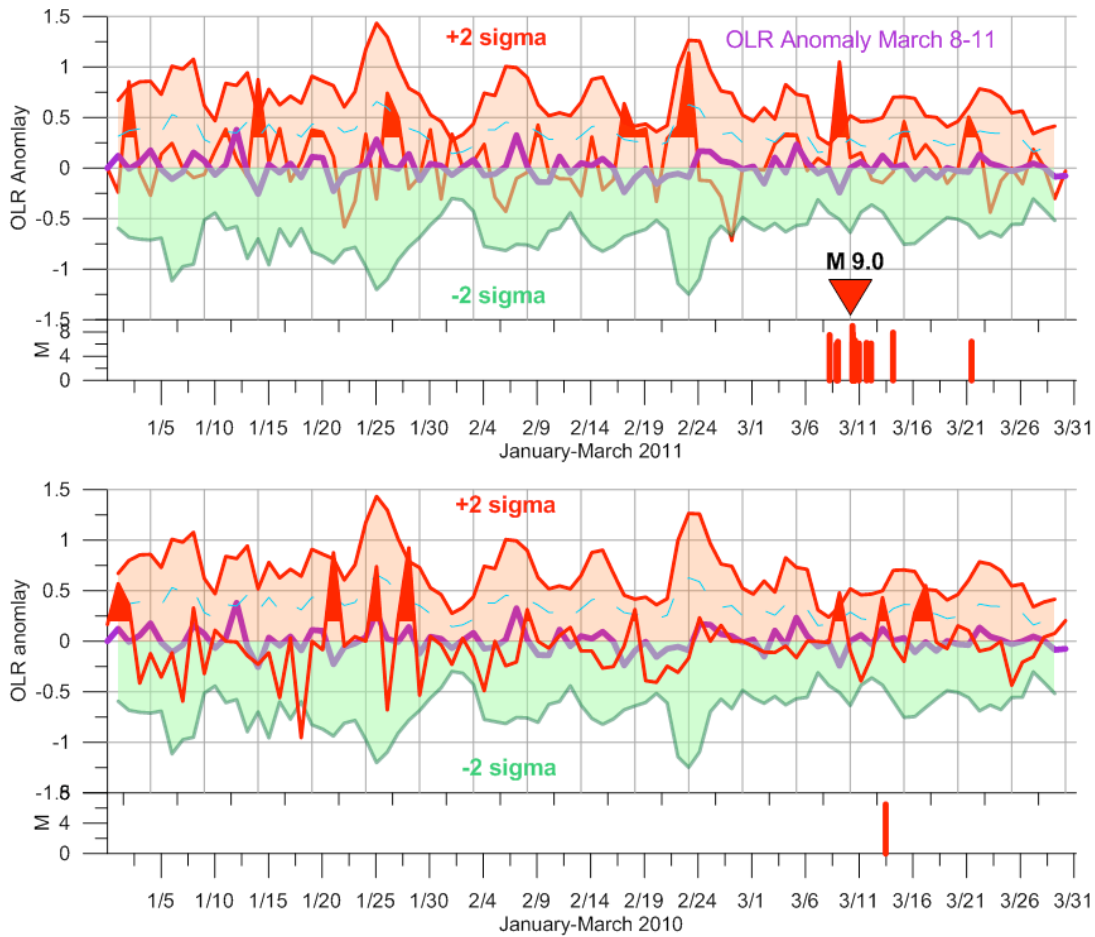

Figure 3. Time series of OLR atmospheric variability observed within a 200 km radius of the Tohoku earthquake (top to bottom). A.) Day-time anomalous OLR from January 1- March 31, 2011 observed from NOAA-15 AVHRR (06.30LT) B.) 2001, seismicity (M>6.0) within 200km radius of the M 9.0 epicenter. C.) Day-time anomalous OLR from January 1- March 31, 2010 observed from NOAA-15 AVHRR (06.30LT) D.) seismicity (M>6.0) within 200km radius of the M 9.0 epicenter for 2010.

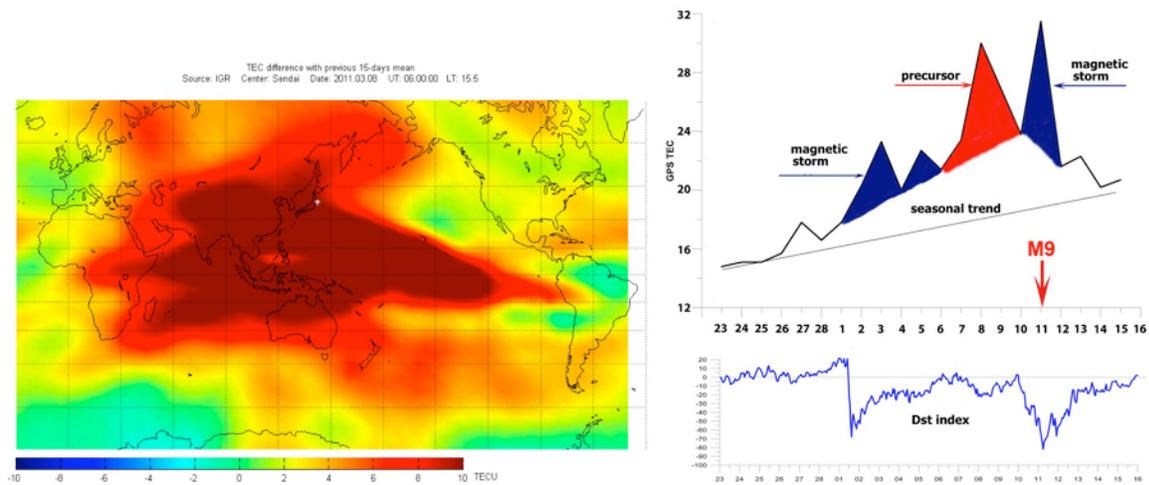

Figure 4. GIM GPS/TEC analysis. (A) Differential TEC Map of March 8, 2011 at 15.5 LT; (B) time series of GPS/TEC variability observed from Feb 23 to March 16, 2011 for the grid point closest to epicenter for the 15.5 LT; and (C) The Dst index for the same period in (B). The Dst data were provided by World Data Center (WDC), Geomagnetism, Kyoto, Japan.

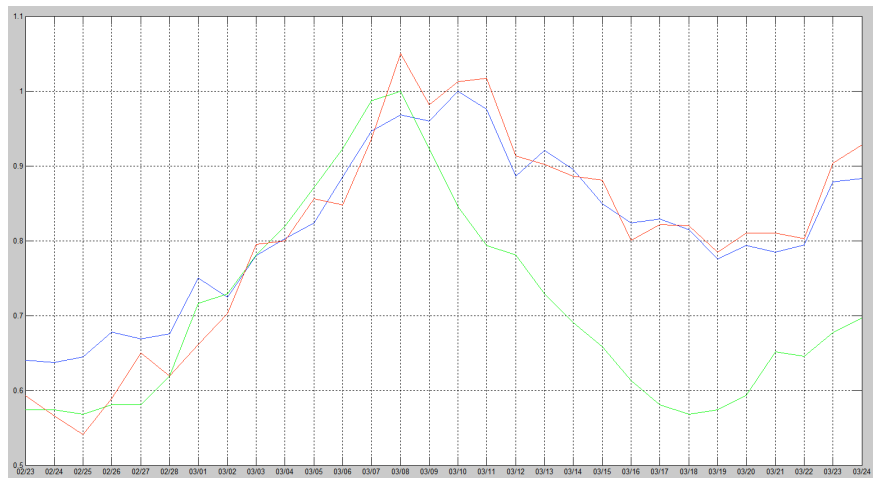

Figure 5 Normalized variations of solar F10.7 radio flux (green), GEC index (blue) and modified GEC (30° around the epicenter of Tohoku earthquake) red.

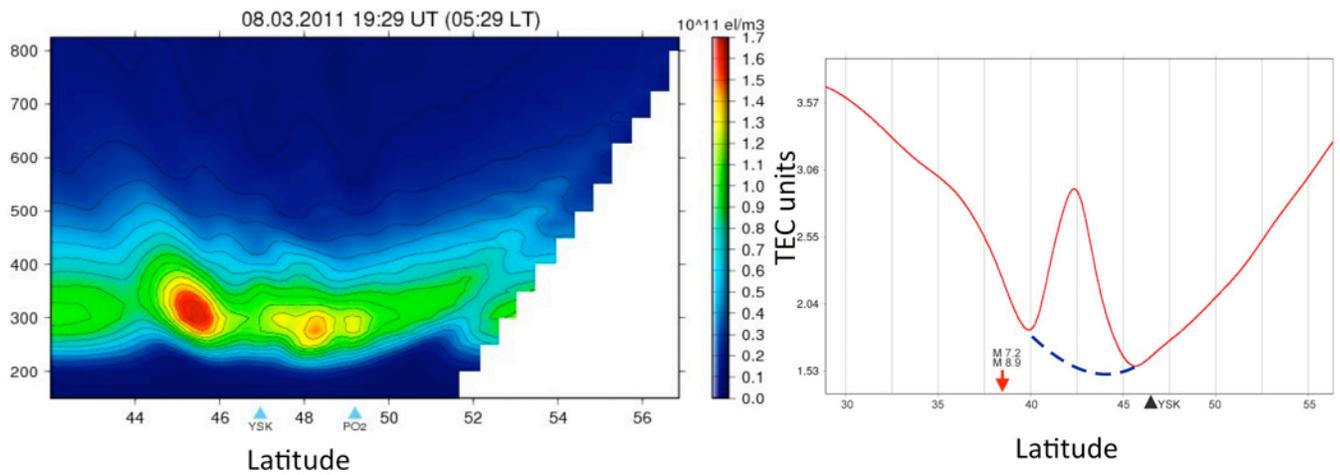

Figure 6. Ionospheric tomography reconstruction over Japan using COSMOS (Russia) satellites and receivers installed at Sakhalin Island (Russia). See Fig.1. (A) Tomography map of March 8, 2011, 05.29 LT; and (B) Ionospheric reconstruction over Japan for March 2011. Blue dash line the TEC reference line (without earthquake influence). Red arrow - location of M9.0 earthquake, and black triangles, location of the ground receiver.

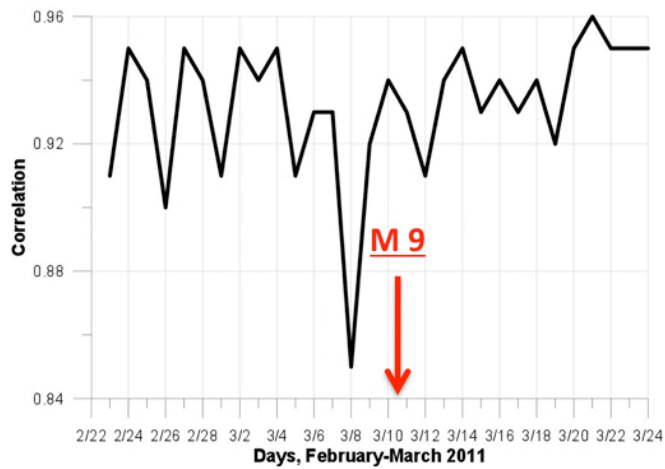

Figure 7. foF2 data cross-correlation coefficient between daily variations at Kokubunji and Yamagawa stations.